# SMLT: A Serverless Framework for Scalable and Adaptive Machine Learning Design and Training

Ahsan Ali[1], Syed Zawad[1], Paarijaat Aditya[2], Istemi Ekin Akkus[2],
Ruichuan Chen[2], Feng Yan[1]
[1]Department of CS, University of Nevada, Reno, NV, USA {aali,szawad,fyan}@nevada.unr.edu
[2]Nokia Bell Labs, Stuttgart, Germany {istemi_ekin.akkus,paarijaat.aditya,ruichuan.chen}@nokia-bell-labs.com

**ABSTRACT**

In today's production machine learning (ML) systems, models are continuously trained, improved, and deployed. ML design and training are becoming a continuous workflow of various tasks that have dynamic resource demands. Serverless computing is an emerging cloud paradigm that provides transparent resource management and scaling for users and has the potential to revolutionize the routine of ML design and training. However, hosting modern ML workflows on existing serverless platforms has non-trivial challenges due to their intrinsic design limitations such as stateless nature, limited communication support across function instances, and limited function execution duration. These limitations result in a lack of an overarching view and adaptation mechanism for training dynamics, and an amplification of existing problems in ML workflows.

To address the above challenges, we propose SMLT, an automated, scalable and adaptive serverless framework to enable efficient and user-centric ML design and training. SMLT employs an automated and adaptive scheduling mechanism to dynamically optimize the deployment and resource scaling for ML tasks during training. SMLT further enables user-centric ML workflow execution by supporting user-specified training deadline and budget limit. In addition, by providing an end-to-end design, SMLT solves the intrinsic problems in serverless platforms such as the communication overhead, limited function execution duration, need for repeated initialization, and also provides explicit fault tolerance for ML training. SMLT is open-sourced and compatible with all major ML frameworks. Our experimental evaluation with large, sophisticated modern ML models demonstrates that SMLT outperforms the state-of-the-art VM-based systems and existing serverless ML training frameworks in both training speed (up to 8×) and monetary cost (up to 3×).

## 1 INTRODUCTION

**Motivation:** The success of machine learning (ML) has prospered intelligent applications, such as translation [18], image classification [30] and autonomous driving [52]. Designing and training ML models have become a daily routine for many ML engineers and practitioners. In production ML systems, models are continuously improved, trained, and deployed for providing inference services. Therefore, today's ML design and training are part of a continuous workflow with different training tasks that have various dynamic resource demands, usually involving pre-processing, architecture engineering, hyperparameter tuning and training via either manual efforts or automated approaches (e.g., neural architecture search [27]).

**Limitation of state-of-art approaches:** Thanks to its elastic resource offering, cloud is one preferred choice to perform ML design and training workloads. In particular, Machine-Learning-as-a-Service (MLaaS) offered by major cloud providers, such as Amazon SageMaker [31], Microsoft Azure [5], and Google Cloud [8], support optimized execution of ML tasks. Despite various cloud management tools, these systems usually require users to have extensive expertise. For instance, Amazon SageMaker provides hundreds of instance types with different computation and communication capabilities under a wide range of prices. Selecting the appropriate instance types for scaling up and the suitable number of instances for scaling out from such a wide selection is a non-trivial task. Recent efforts [37, 41, 59] on the optimization of ML task deployment target the automation of the deployment process. However, they incur significant monetary costs just for tuning the environment before training (up to 60% of the total) [59]. These approaches are not ideal for the state-of-the-art ML workloads that exhibit highly dynamic resource demands throughout the training process, such as NAS [27], dynamic batching [23] and online training.

**Key insights and contributions:** Serverless computing is an emerging cloud computing paradigm with one popular incarnation being Function-as-a-Service, where the unit of computation is a function.[1] Serverless computing has great potential to match the requirements of ML workloads with dynamic resource demands, because it aims to eliminate the resource and scaling management from users and offers a flexible "pay-as-you-go" pricing model. As such, serverless computing is gaining popularity for inference from both academia [17, 36, 40] and industry, with the latest example being the AWS SageMaker Serverless Inference [1, 3, 12].

On the other hand, ML model design and training on serverless platforms is still at an incipient stage. Initial attempts to enable ML training on serverless platforms [22, 28] are limited to specialized frameworks, support only simple approaches or small models (e.g., linear regression, MNIST), and scale poorly to larger and more sophisticated ML training tasks. Due to these limitations, the advantages in performance and cost when using serverless computing for large practical training tasks have been questioned [33].

Alleviating these concerns requires three challenges to be addressed. *First*, modern ML workflows usually have dynamic scaling and resource demands. For example, when the batch size [23] or model size changes during the training process, it may inevitably change the scalability and resource demands of the training tasks. Not adjusting the underlying serverless resources accordingly may result in lower performance and/or cost overheads. Hence, leveraging a generic serverless platform for ML training tasks requires overarching monitoring of the training dynamics, and a resource adaptation mechanism. The stateless nature of serverless platforms creates a challenge for such a mechanism.

---

[1]Other incarnations of serverless computing include, e.g., the Container-as-a-Service with a full automation capability.

*Second*, existing serverless platforms either do not provide user-centric deployment guarantees (such as meeting training deadlines and training budgets) or incur a heavy runtime overhead in doing so [22]. To achieve these goals and to optimize the cost and resource usage, the number of serverless functions and their memory configurations may need to be adjusted at a fine grained level for different ML tasks. These adjustments require going beyond simply monitoring the progress of the training pipeline, and introducing an optimizer that can, at runtime, automatically search for necessary configurations to meet a user-specified goal.

*Finally*, the end-to-end ML training platform based on serverless computing needs to address the challenges arising with the increasing size and sophistication of the ML models as well as the different training frameworks that can be used. It is already challenging to scale ML training using the traditional infrastructure with dedicated computation and network resources, and using a serverless platform may amplify any existing issues. For example, the communication overhead for model synchronization already presents challenges; using stateless serverless functions that communicate via external storage can exacerbate these bottlenecks. In addition, the stateless functions may experience substantial repeated framework initialization overheads.

To address the above challenges, we propose SMLT — a serverless framework for scalable and adaptive ML design and training. To enable an overarching view of the ML workflow and the swift scale-adaptation, we propose an automated and adaptive scheduler for deploying and scaling training instances dynamically on the fly. Our scheduler can cater to varying resource requirements during the execution of ML tasks, such as the dynamic batch size scheduler [23] and the architecture exploration in NAS [27]. To achieve the user-centric deployment and execution guarantees, we offer a light-weight Bayesian optimizer to automate and optimize the ML task deployment and resource scaling to meet the user-specified performance and cost goals.

SMLT provides an end-to-end ML training service. It addresses the challenges caused by the stateless nature of serverless functions, including the initialization and communication overheads. Specifically, it employs two key system components: 1) a hybrid storage consisting of a cloud object store for infrequently-accessed training data and an in-memory key-value store for frequently-accessed model parameters, and 2) a hierarchical model synchronization scheme exploiting parallelism during model synchronization. In addition, SMLT is designed to be agnostic of ML frameworks by abstracting the implementations of their common interfaces (e.g., input pipelines, gradient transfers).

**Experimental methodology and artifact availability:** To our knowledge, SMLT is the *first* fully-automated ML design and training framework that can support modern ML workflows in a scalable and cost-effective manner with the emerging serverless principles. To demonstrate its flexibility, we evaluate SMLT using three different ML frameworks (i.e., Tensorflow [14], Pytorch [45] and MXNet [24]), and show that our results are consistent across all of them. We perform our evaluation across a variety of ML workflows, including user-specified goals to minimize training time or budget, dynamic batching, online learning and NAS, and show that SMLT can reduce cost by up to 3x compared to the state-of-the-art. We also demonstrate that our system scales well with an increasing number of workers, leading to an overall reduction of up to 8x in the total training time when combined with other optimizations. We open-source the codebase of SMLT for public access[2].

**Limitations of the proposed approach:** One key challenge of SMLT is the unavailability of GPUs in typical public serverless platforms. However, we believe with the wide adoption of serverless computing in academia and industry for a wide range of applications, public cloud providers will soon offer GPUs for serverless computing as well. Initial prototypes of serverless platforms with GPU support have already started emerging [49], and ML training using GPUs in a serverless environment has already been employed in HPC [39] with promising results. We believe SMLT will demonstrate benefits on GPUs similar to our CPU evaluation.

## 2 BACKGROUND AND MOTIVATION

In this section, we give an overview of the latest ML workflows and how these workflows utilize cloud resources. We further motivate our work by discussing the state-of-the-practice and state-of-the-art cloud ML systems and identifying their limitations.

### 2.1 Modern Machine Learning Workflows

**Distributed Machine Learning.** Distributed data parallel training is typically employed to train modern ML models, which can contain up to billions of learning parameters [59], and require terabytes of training data [20]. Each worker node trains the model on a partition of the training data, and synchronizes weight updates frequently with other nodes. The choice of the synchronization approach can have a big impact on the training speed. In practice, Parameter Servers (PS) [7, 15, 38] and Ring-allreduce [19, 20, 24, 51] are typically used to coordinate the synchronization of weights across all worker nodes.

**Machine Learning with Dynamic Resource Requirements.** The life cycle of an ML model development includes pre-processing the training data, designing an appropriate model, tuning hyper-parameters, and training. The resource requirements across these phases as well as within each phase can vary significantly and often require dynamic resource provisioning. For example, during the model design, engineers usually employ the autonomic design technique known as Neural Architecture Search (NAS) [26, 27] to find an optimized model architecture. During this phase, different architectures are crafted, trained, and evaluated according to the evaluation results of the previous trials, resulting in varying resource requirements for the new candidate model. Unless these new requirements are met by scaling the underlying memory and communication resources, the optimal execution cannot be achieved in terms of throughput and monetary cost.

Similarly, a model training task often employs dynamic batching [61] to improve the training speed and model accuracy, such that the batch size can change between iterations or epochs during training. As a result, memory requirements may need to be adjusted. In addition, a larger batch size presents opportunities for intra- and inter-node parallelization, such that having more workers can benefit the training process. For these reasons, modern ML workflows often have dynamic resource demands and require effective auto-scaling for optimal execution. SMLT is designed to handle such dynamic resource requirements, and we demonstrate its effectiveness after applying it to NAS and dynamic batching systems in Section 5.

---
[2]Link removed for double-blind review.

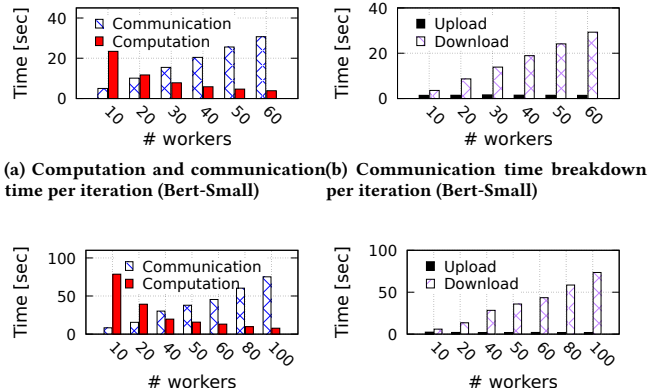

Figure 1: Scalability of Bert-Small and Bert-Medium using Siren [56].

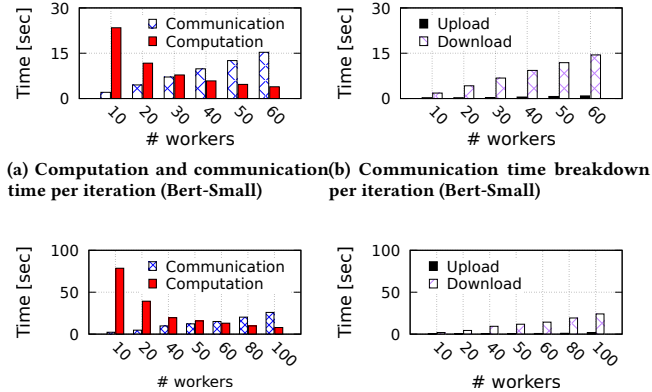

Figure 2: Scalability of Bert-Small and Bert-Medium using Cirrus [22].

## 2.2 Machine Learning on the Cloud

Many existing ML systems for training large-scale models require a significant amount of computational resources that may not be available in a user's own infrastructure. Cloud computing and MLaaS systems can address this problem [47]. Hereafter, we briefly describe some well-known solutions.

**Infrastructure-as-a-Service (IaaS).** Cloud providers offer virtual machines (VMs) with a wide range of computational capabilities. This scheme is known as Infrastructure-as-a-Service (IaaS). IaaS provides on-demand resources with a time-based cost model. In an IaaS environment, users are fully responsible for deploying, managing, and provisioning the compute resources based on their requirements. Although it is a flexible option for practitioners, the extensive resource management overhead and limited scaling support make IaaS a less ideal candidate for training modern ML models, especially for ML practitioners with limited system expertise.

**Machine-Learning-as-a-Service (MLaaS).** To address the challenges of running ML tasks in an IaaS environment, cloud providers offer Machine-Learning-as-a-Service (MLaaS) to provide additional support for ML tasks. Many commercial MLaaS systems are still based on a pool of VMs with varying computational and communication capabilities, require a shared storage for training data, and offer some degree of scalability. Although presenting a plethora of options to choose from [6, 8, 13], these platforms require users to manually configure the required resources. Furthermore, to cover dynamic resource requirements during ML tasks and provide robustness against failures (e.g., out of memory), they typically require users to over-provision the configured resources, incurring inefficient resource utilization and high monetary cost. These problems are exacerbated when the ML workflows include continuous learning and training with continuously incoming training data. As a result, ML practitioners still lack a fully-automated MLaaS solution to achieve dynamic resource allocation for various ML workflow phases with both performance and cost optimizations.

**Machine Learning with Serverless Computing.** Function-as-a-Service (FaaS), along with the Container-as-a-Service (CaaS) with a full-automation capability, are widely-realized forms of serverless computing and have been gaining popularity in both industry and academia. The elastic nature and the fine-grained resource management make these serverless realizations attractive to many application developers, including ML design and training developers. Besides removing the burden of resource management and scaling from ML developers, they also offer a flexible billing model and the continuously-improved execution startup delays [16, 42, 43].

In fact, FaaS has been actively studied for ML model serving [17, 19] and ML training [21, 22, 28, 33, 56]. Although it is easy to see that model serving can benefit from FaaS' advantages (e.g., scalability, billing flexibility), the benefits for ML training are still not clear. Cirrus [22] is a specialized ML framework that addresses the resource limitations (e.g., memory, storage) in FaaS environments and applies it to linear regression. Siren [56] showcases that cost efficiency and training speed can be improved by adjusting the number of workers during different phases of training via reinforcement learning and using the MNIST model with a maximum of 10 workers.

Although these frameworks demonstrate that FaaS can be a viable alternative for ML training, their evaluation with small models and limited resources raises questions about the feasibility of training large-scale models with FaaS and serverless computing. It would not be surprising that increasing the number of training workers to process large ML models will significantly increase the communication overhead, the training time and monetary cost. For example, in a recent study, Jiang et al. [33] compare IaaS and FaaS platforms, and highlight issues in serverless settings that affect the performance and cost of ML training.

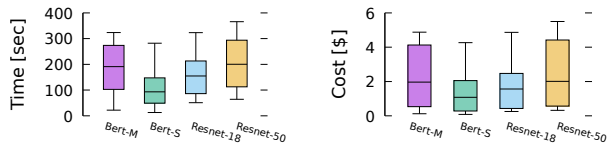

(a) Per iteration computation time distribution　(b) Per iteration computation cost distribution

**Figure 3: Per iteration computation time and cost distributions with varying deployment configurations for Bert-Medium, Bert-Small, Resnet-18 (Tensorflow), as well as Resnet-50 (MXNet).**

To confirm these questions, we evaluated two large Natural Language Processing (NLP) models (i.e., BERT-small [48] and BERT-medium [55]) using these two approaches.[3] Figure 1 shows that the computation time decreases with increasing number of workers in Siren [56], but so does the communication time due to the use of S3 for storing intermediate parameter updates. With more than 20-40 workers, the total training time increases due to the communication overhead. Figure 2 shows a similar behavior for Cirrus [22]. These experiments show that communication can be a major bottleneck for training complex ML models in a serverless environment.

We further evaluated the performance of four models with varying number of workers between 10 to 200 and with three different sizes of memory allocation to these workers (3 GB, 6 GB and 10 GB). Figure 3a and 3b show the training time distribution and the cost distribution per iteration, respectively. These results highlight that an incorrect selection of workers and inefficient resource allocation to them can have significant impacts on training time and cost in a serverless environment. Furthermore, the high variance in these values shows that it is not trivial to find the 'right' values for these parameters. Cirrus [22], Siren [56] and LambdaML [33], however, assume that the users know these values *before* running the training. In other words, these systems do not support dynamic resource allocation during training, leading to sub-optimal training time and causing higher monetary costs. In summary, these systems lack support for the needs of modern ML workflows with large models, dynamic resource requirements and user-centric training goals.

## 3 SMLT DESIGN

In lieu of the aforementioned shortcomings to support dynamic resource requirements, user-centric deployment goals and large-scale ML models, we propose SMLT. In this section, we explain the rationale behind the SMLT design, present our goals, and then describe the building blocks to achieve them.

Our general aim is to design a generic framework that provides the easy-to-use functionality for ML practitioners to design and train their ML models efficiently and economically on a serverless platform. The framework should allow ML practitioners to focus on business logic while at the same time truly realizing the transparency and scalability promises of serverless computing. In particular, we design our building blocks to target three main goals.

---
[3]Note that Siren's source code is not available, and Cirrus does not support the ML frameworks for these models. Nevertheless, we replicated these approaches to the best of our abilities.

- We aim to offer an overarching view of dynamic ML workflows to enable adaptive and efficient serverless scaling.
- We aim to support user-centric deployment goals (e.g., training deadlines and budget limits) for running ML workflows on serverless platforms.
- We aim to achieve scalability by enabling the handling of large and sophisticated ML workflows while abstracting out the limitations of the underlying serverless platforms.

### 3.1 Overarching View and Dynamic Adaptation

The underlying primitives in existing serverless platforms can support individual ML training tasks well. However, for modern dynamic ML workflows, such as dynamic batching [23] and NAS [27], today's serverless systems often fall short. The fundamental reason is the lack of a component that maintains an overarching view of the training dynamics of ML tasks. For example, when the batch size or model size changes over the training process, it may inevitably change the scalability and resource demands of the training task. Not adjusting the underlying serverless resources accordingly may cause performance and/or cost issues. The stateless nature of serverless platforms, and hence their inability to carry over the training dynamics across function invocations, creates a challenge for supporting such an adaptation mechanism.

To solve this challenge, we propose a *training dynamics aware* design to enable an overarching view of ML workflows. To keep track of the training progress, we propose a monitoring component (called *Task Scheduler* in Section 4.1) to collect the training information (e.g., time taken to complete one iteration, batch size changes), and maintain it across function invocations. The task scheduler continuously monitors for changes in training information, and upon detecting change, activates an optimizer to determine the new resource allocation to meet the optimization targets. The scheduler then allocates resources according to the newly optimized resource decisions, such that the training can continue with new resources (e.g., number of workers, memory configuration).

### 3.2 User-centric Deployment and Execution

To guide the above process, we employ a user-centric deployment and execution approach. ML practitioners often have different goals in ML model training, such as meeting deadlines or staying within a monetary budget. These desired goals are not supported by today's MLaaS platforms (see Section 2). The serverless computing paradigm creates an unprecedented opportunity to provide these goals because of its advantages in scaling and resource management. By embracing serverless computing, SMLT alleviates the burden of reserving and scheduling resources from ML practitioners, and allows them to focus on their ML model development, while guaranteeing their time and budget constraints are met.

**Definitions and Terminology.** Our approach takes a user's requirements as input to optimize the deployment and execution of ML tasks. Without loss of generality, we use dynamic batching as an example to illustrate our approach. Let $\mathcal{B} = \{b_1, b_2, \cdots, b_n\}$ denote the batch scheduler, where $b_i$ denotes the batch size at the $i^{th}$ epoch. Let $C = \{c_1, c_2, \cdots, c_n\}$ denote the system configuration across epochs as a tuple of scale-out factor (number of workers $c_i^{\text{worker}}$) and scale-up factor (memory size $c_i^{\text{memory}}$), i.e., $c_i = \langle c_i^{\text{worker}}, c_i^{\text{memory}} \rangle$. Let $T_\mathcal{B}(C)$ denote the training time using configuration $C$ for a given batch scheduler $\mathcal{B}$, and $S_\mathcal{B}(C)$ denote the monetary cost using configuration $C$ for a given batch scheduler $\mathcal{B}$. In addition, let $T_{max}$ denote the user-specified training deadline and $S_{max}$ denote the user-specified monetary budget for training.
**Example Scenario.** We give an example scenario where a user wants to meet the training deadline while minimizing the cost. This scenario can be modeled as an optimization problem as follows:

$$\begin{aligned} \text{minimize} \quad & S_\mathcal{B}(C) \\ \text{subject to} \quad & T_\mathcal{B}(C) \leq T_{max}. \end{aligned} \quad (1)$$

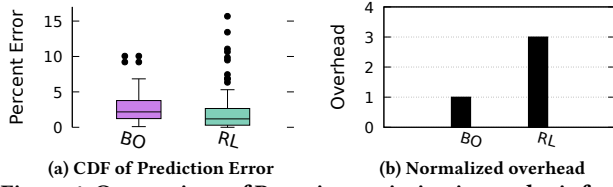

(a) CDF of Prediction Error    (b) Normalized overhead

**Figure 4: Comparison of Bayesian optimization and reinforcement learning in terms of accuracy and training overhead.**

To obtain the training cost $S_{\mathcal{B}_j}(C_i)$ of a particular deployment configuration $C_i$ under a given batch scheduler $\mathcal{B}_j$, we need to first profile the training time per iteration (including the training throughput and communication time), and then use the cloud cost model to calculate the monetary cost. However, to exhaustively search the training cost of all possible deployment configurations to identify the minimum cost is prohibitively expensive as we need $n(C)$ number of profiling experiments. Therefore, we need an efficient optimization algorithm to explore the search space to reduce the optimization overhead. While there are plenty of optimization algorithms in the literature which could potentially be adopted here, we propose a customized Bayesian optimizer, inspired by [59], to efficiently solve this optimization problem.

There are other example scenarios with user-centric goals. For example, users may want to minimize training time with a budget, or users may simply want to finish training as fast as possible.
**Bayesian Optimization.** There are a few classic optimizers (e.g., Bayesian optimization and reinforcement learning) to solve the aforementioned optimization problems. We compare their performance in terms of accuracy and training overhead in Figure 4. It can be observed that, for the same prediction accuracy demonstrated in Figure 4a, reinforcement learning incurs 3× overhead compared to Bayesian optimization. In light of this result, we choose Bayesian optimization over reinforcement learning [22] to speed up the optimization process because it is much lighter-weight, and importantly, does not require additional training. This makes it more suitable for our goals in improving training speed and increasing cost-effectiveness.

The Bayesian Optimizer searches for the optimal deployment by profiling the throughput of the system under randomly chosen configurations. Based on the profiled throughput results, the Bayesian optimizer estimates the most beneficial configuration to profile next. This search is done in an iterative manner until the expected improvement is small enough or the predefined max number of iterations is reached, thus finding the optimal configuration.

In doing so, SMLT exploits the serverless paradigm and its 'pay-as-you-go' model with fine-grained resource allocations. SMLT's approach differs from the approach described in [59], whereby the Bayesian algorithm can run only once before the start of the ML task due to its high monetary cost associated with profiling in a VM-based cloud infrastructure, thus leaving little budget for the actual training tasks.

In its iterative configuration search, SMLT uses a two-dimensional search space where the worker memory size varies from 128 MB to 10 GB with an increment of 1 MB [11] and the number of workers varies depending on the model size and training parameters. The cost function for a particular deployment configuration is set as $C_i$, and this cost function is specified as part of user requirements like the example scenarios aforementioned.

We employ the widely-used Gaussian Process Regression [29] to calculate the posterior distribution. For the acquisition function, we use the Estimation Improvement (EI) [44] since it requires no hyperparameter tuning. Specifically, our EI is defined as: $EI(C_i) = (y_{max} - \mu(C_i))\beta(\gamma(C_i)) + \delta(C_i)\theta(\gamma(C_i))$, where $C_i$ is a deployment configuration with $m$ memory sizes and $n$ number of workers, and $y_{max}$ is the current lowest value from all explored tuples. $\mu$ and $\delta$ are the predictive mean and standard deviation functions, with the $\beta$ and $\theta$ being the predictive cumulative distribution function of standard normal and the probability density function of standard normal, respectively.

Based on the output of the Bayesian optimizer, our task scheduler deploys the training task accordingly with the optimized number of workers (scale-out) and the optimized worker memory configuration (scale-up). It is worth noting that, different stages of a training task may have different performance and resource requirements (Section 2). Therefore, our optimizer is designed to consider these dynamic resource requirements during training to improve the overall performance and optimize the resource usage during training. These optimizations are made possible by our user-centric deployment module and the on-demand scalability, as well as the pay-as-you-go billing model provided by serverless computing; thus, enabling us to provide cost-efficient solutions that honor user-specified training deadline and budget requirements.

Altogether, SMLT provides significant advantages over existing MLaaS platforms and other serverless ML training frameworks, and provides user-centric deployments and executions in a fully-automated, scalable, and adaptive fashion. As a result, SMLT not only improves training performance in a serverless environment while meeting user requirements, but more importantly, allows users to focus on the design and logic of ML algorithms, thus enhancing their productivity.

### 3.3 End-to-end Scalability

Using serverless primitives to design an ML training platform presents unique challenges for scalability. Here, we briefly describe these challenges and our approaches to address them. We will elaborate upon the end-to-end design of SMLT in Section 4.
**Challenge 1: Communication Overhead.** Two unique characteristics of a serverless platform are the stateless nature of a serverless function and the limited communication performance across functions. These characteristics force serverless ML platform designers to utilize an external storage for model synchronization. Hence, the communication load/pattern generated by the chosen model synchronization scheme as well as the choice of the storage service play a crucial role. If not chosen properly, the communication overhead can easily overshadow the gains achieved via splitting computation across a large number of serverless workers, as demonstrated in Figures 1 and 2.

**Approach: Hybrid Storage Enabled Hierarchical Model Synchronization.** We observed from our experimentation that a hybrid storage service combined with a hierarchical synchronization mechanism (similar to ScatterReduce suggested by LambdaML [33]) scales well for large ML models.

There is a large amount of data generated and used in ML tasks. Such data falls broadly into two types according to their sensitivity to latency. 1) The access to the first type of data is latency sensitive, e.g., the metadata containing the gradient-worker mapping information, the gradients produced during each training iteration, etc. For this type of data, we employ a fast storage medium to satisfy the latency-sensitive demands of the synchronization scheme. In particular, an in-memory key-value store (e.g., Redis) can be used as a *parameter store* to store this type of data. 2) The access to the second type of data is much more infrequent. For example, the training code and dataset are accessed only few times during an epoch or after every epoch. To strike a balance between performance and cost, we use a cloud-based object store (e.g., AWS S3). We present the details of our hybrid storage in Section 4.3.

Figure 5 presents our hierarchical synchronization mechanism. After each training iteration, the hierarchical synchronization mechanism takes the model gradients generated by each worker as input. The *shard generator*, residing in each of the $n$ workers, divides the input gradients into $m$ equal-sized shards ❶. These shards are uploaded to the *parameter store* which acts as a communication intermediary between the stateless serverless workers ❷. Each serverless worker also acts as a shard aggregator. Each shard aggregator is responsible for downloading and aggregating its assigned shards generated by all workers ❸. For simplicity, we assume that $n$ equals $m$.[4] As a result, the 'shard aggregator 1' in Figure 5 is responsible for aggregating the first shard from all workers to perform a mean operation. The resulting value, *aggregated shard*, is then uploaded to the parameter store by each shard aggregator ❹. Finally, the *global aggregator* residing in each worker downloads all the aggregated shards, and reconstructs the updated model for the next iteration ❺. We describe in detail when and how this scheme is used in Section 4.2.

**Challenge 2: Serverless Platform Quirks.** Serverless platforms typically limit the maximum duration of each executing function (e.g., 15 minutes in AWS Lambda [2]). Getting around this limitation as well as recovering from faults during function execution require keeping track of function states at regular intervals and restoring state during a restart. In addition, each function invocation may have non-trivial initialization overheads, making it important to minimize the number of restarts a function encounters. For example, initialization overheads could arise due to model loading within the function code itself. Other overheads include anomalous serverless platform behavior (which we encountered during our experiments on AWS Lambda), such as high invocation delays while invoking functions asynchronously and while invoking via the 'Map' state in AWS Step Functions [4]. Section 4.1 presents additional details.

---
[4]If $m$ is greater than $n$, then each worker is responsible for aggregating multiple shards. Choosing $m$ less than $n$ will cause some workers to be idle during aggregation, which will affect performance.

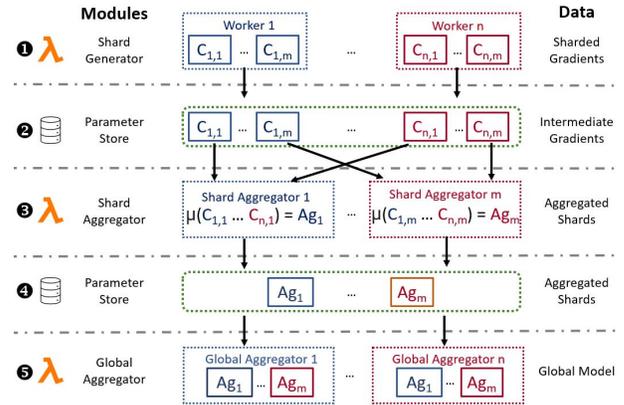

**Figure 5: Hierarchical model synchronization mechanism.**

**Approach: Task Scheduler.** This challenge motivates us to create a coordination mechanism to provide the maximum flexibility and serverless platform independence, while invoking and monitoring function instances. To enable this coordination, we use a component called *task scheduler* in our approach. The task scheduler monitors the progress of each function instance, enables function instances to run for the maximum allowed execution duration, and performs the checkpointing for restarting functions (either due to faults or due to execution duration limits). We detail the task scheduler in Section 4.1.

## 4  SMLT FRAMEWORK

In this section, we describe the details of the SMLT framework. Broadly speaking, SMLT adopts an architecture similar to a parameter server based distributed training using a fleet of stateless serverless functions. Although some existing frameworks [22, 33, 56] use a similar architecture, SMLT additionally offers the capability of intelligently and dynamically determining resource allocation (i.e., number of workers, memory sizes) that enables efficient and user-centric training of the state-of-the-art ML models with billions of learning parameters.

Table 1 gives an overview of SMLT, which consists of three main system modules. The *end client* is responsible for code deployment, resource allocation based on user-centric deployment and execution configurations, and ML task scheduling. The *workers* are mainly responsible for training and updating the global model. In *storage*, following our hybrid approach, the parameter store maintains the frequently-accessed intermediate model updates between workers, and the object store maintains the infrequently-accessed training code and data. Figure 6 shows the interactions between these different modules.

### 4.1  End Client

The end client provides an interface for users to interact with the cloud infrastructure. In particular, the end client is responsible for three tasks: 1) ML model and code deployment, 2) resource management which configures a deployment (e.g., number of workers, per-worker memory size) based on user-centric requirements and optimizations, and 3) scheduling tasks during training. Each user has her own end client, which can be deployed either on a regular VM in the cloud or on a user's local machine.

Table 1: Modules of SMLT and their functionalities.

| Module | Submodule | Functionality |
|---|---|---|
| ① End Client (§4.1) | ⓐ Artifact Manager | Uploads user's training code and training data to the cloud-based object store ③ⓐ. |
| | ⓑ Resource Manager | Allocates resources for training workers ② based on user-centric requirements (e.g., number of workers, and memory per worker). |
| | ⓒ Task Scheduler | Invokes workers ② for training, keeps track of training progress, and restarts failed workers when necessary. |
| ② Worker (§4.2) | ②ⓐ Data Iterator | Loads training data to the local storage of the function from the object store ③ⓐ. |
| | ②ⓑ Minibatch Buffer | Keeps track of the training data for every iteration used by the trainer ②ⓒ. |
| | ②ⓒ Trainer | Runs the user-defined training code using the training data from the minibatch buffer ②ⓑ. |
| | ②ⓓ Hierarchical Aggregator | Updates user model parameters after every iteration through the parameter store ③ⓑ. |
| ③ Storage (§4.3) | ③ⓐ Object Store | Stores training code and training data. |
| | ③ⓑ Parameter Store | Stores intermediate training parameters of all workers ② after every iteration. |

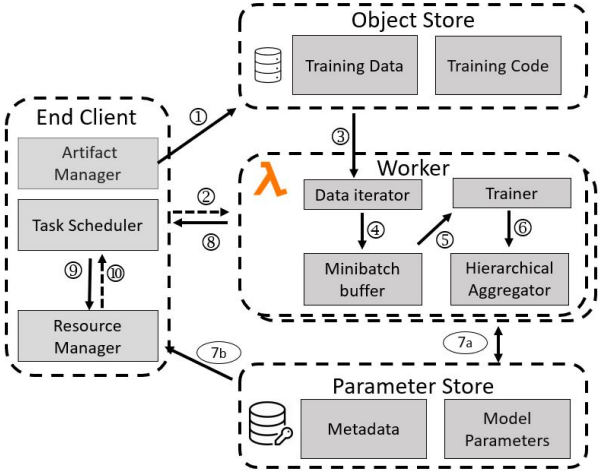

Figure 6: Workflow of SMLT. Solid lines indicate data transfer, and dashed lines indicate control signals.

The end client has three submodules: artifact manager, resource manager, and task scheduler.

**Artifact Manager.** The artifact manager is responsible for packaging the training code and data supplied by the user as input to SMLT, and uploading them to the object store (Step ① in Figure 6).

**Resource Manager.** The resource manager uses the training configuration (e.g., batch size) along with user-centric deployment requirements as input, and will dynamically adapt the resources for optimal training (Step ⑨).[5] Specifically, the resource manager controls the number of workers involved in a training phase, as well as the memory allocated to each worker. The appropriate values in different phases of the training are provided by the Bayesian optimizer according to user-centric goals and the training progress (see Section 3.2), and then shared with the task scheduler (Step ⑩). The user can also override these values manually. Note that, commercial serverless platforms typically provide a single control parameter for resources. For example, in AWS Lambda, other resources (e.g., number of computing cores, network bandwidth) are proportionally assigned by the allocated memory of a function instance.

The optimal training depends on the user's training requirements, training dataset, model size, and the cloud infrastructure's profile. SMLT is capable of optimizing the resources during the hyper-parameter search, dynamic batching and online learning for speed and cost-effectiveness. The optimization and adaptation of resources are triggered every time there is a change in training configuration obtained from the worker output (e.g., batch size or hyper-parameter change).

**Task Scheduler.** After the resources are determined by the resource manager, the task scheduler starts the training workers (Step ② in Figure 6) and will track their progress (Step ⑧). When the workers finish an iteration, their results are collected and fed back to the resource manager to be used in the Bayesian optimizer (Steps ⑧ and ⑨).

Tracking the progress of the workers by a separate task scheduler is important for three reasons: First, due to the stateless nature of the serverless workers, the used ML model needs to be obtained, and the ML framework used in the training code needs to be initialized by the worker. This initialization overhead depends on the model size and the specific ML framework, and can take multiple seconds (e.g., 4 seconds for Resnet-18 on Tensorflow) that quickly accumulate with function restarts. Therefore, a separate task scheduler can maximize the utility of each training worker by running it close to the limit of the function execution duration enforced by the underlying serverless platform (e.g., 15 minutes on AWS Lambda) and across multiple iterations; therefore, the initialization overheads are amortized. After a worker is stopped, the task scheduler ensures that a new one is started and continues from the last iteration checkpoint.

The second reason for tracking the progress of each worker by a task scheduler is fault tolerance. A worker failure can derail the progress of the training. The task scheduler, with the help of the resource manager, detects failures by observing each worker's output (Step ⑦ⓑ): The resource manager checks the output of each worker to determine whether a failure occurred during the training iteration. A successful update of the gradient parameters by a worker sets a flag in the output. The lack of this flag signals a worker failure, causing the task scheduler to restart it.

---
[5]Except for the initial configuration, where the training resources are arbitrarily assigned within the allowed bounds.

Finally, there may be hidden overheads specific to the underlying serverless platforms. For example, when we experimented with serverless functions invoking other serverless functions asynchronously in AWS Lambda as utilized in LambdaML [33], we observed undocumented invocation delays for the new functions. Another example of such a hidden overhead is the limit on the concurrency of a function within the 'Map' state of a state machine in AWS Step Functions, even when the concurrency is set to be 'infinite' [10].[6] A separate and independent task scheduler in SMLT overcomes these limitations by treating every function invocation as a separate invocation and tracking them accordingly, and forcing the serverless platform to do the same.

### 4.2 Serverless Worker

The main responsibility of each SMLT worker is threefold: 1) obtain the training data, 2) run the training code, and 3) update the model. Each worker consists of the following four submodules that handle these tasks.

**Data Iterator.** Due to the limited storage space within serverless functions, similar to the existing work [22, 56], SMLT stores the training data in external cloud storage like AWS S3 or EFS. The data iterator submodule within each worker is responsible for fetching the appropriate subset of the training data from the external storage at the beginning of every training epoch and storing it within the worker's local disk storage (Step ③ in Figure 6). Furthermore, the data iterator also tracks which training data points have been processed by a worker within an epoch, in case the worker needs to resume training after a restart (due to either a failure or an execution time limit).

**Minibatch Buffer.** The minibatch buffer is responsible for loading a minibatch of training data from the local storage to the memory of the worker during each training iteration (Step ④). The minibatch size varies based on the number of workers and the global batch size during the training.

**Trainer:** The trainer submodule runs the user-defined training code over the minibatch of the training data (Step ⑤). The training code carries out the forward and backward passes to generate the gradients for the current iteration. These gradients are then aggregated with the gradients from other workers via SMLT's hierarchical model synchronization (Step ⑥).

**Hierarchical Aggregator.** The hierarchical aggregator submodule at each worker is responsible for synchronizing the gradients among different workers. As described in Section 3.3, the hierarchical aggregator at each worker first divides the gradients produced by the worker into shards, and uploads them to the parameter store (e.g., Redis). Each worker downloads the shards it is responsible for from the parameter store, aggregates them into a single shard, and uploads the aggregated shard back to the parameter store. Finally, the aggregator at each worker downloads all aggregated shards and produces the updated model for the next iteration (Step ⑦a).

---

[6]Since the writing of this paper, the documentation has been updated to reflect that it is not guaranteed to achieve a requested concurrency.

### 4.3 Hybrid Storage

The choice of the storage system plays a crucial role because it is the main communication mechanism among serverless workers. In SMLT, we employ a hybrid approach that uses a cloud object store, e.g., AWS S3, for infrequently accessed data (few times or after every epoch), and an in-memory key-value store, e.g., Redis, for frequently accessed data (after each iteration). This hybrid approach matches the training phase to a cost-effective object store for bulk data access, while at the same time matching the model update phase to a fast in-memory key-value store for frequent data access.

**Object Store.** The object store hosts the training code and training data provided by the user and uploaded by the end client. The training code uploaded by the end client is accessed a few times at most (e.g., at the beginning of training, or after restarts from failures). Similarly, training data, which is accessed relatively infrequently after every epoch, is also hosted in the object store. Using the object store provides elasticity and cost efficiency for storing such types of infrequently-accessed data.

**Parameter Store.** In contrast, workers access gradient data much more frequently (i.e., after every iteration). During the hierarchical model synchronization, the speed of access to the updated and sharded gradients becomes critical for training speed. We use an in-memory key-value store (e.g., Redis) for this purpose. To avoid the extra cost of running the parameter store unnecessarily during the entire training, we use the light-weight containers from a cloud service (e.g., AWS Fargate, ECS), and keep them only alive during the model synchronization phase.

## 5 EVALUATION

In this section, we evaluate SMLT's capability to enable scalable and adaptive ML workflows via serverless computing. We first demonstrate the effectiveness of the hierarchical model synchronization. Next, we present two different user-centric deployment scenarios. We then validate the performance of SMLT for dynamic batching and online learning. Finally, we show how SMLT can alleviate the burden of dynamic resource provisioning from ML engineers and practitioners during model design and building via an autonomic technique like neural architecture search (NAS).

### 5.1 Experimental Setup

We evaluate SMLT using three popular ML training frameworks: Tensorflow 2.0 [14], MXNet 1.7.0 [24] with gluon-cv 1.4.0, and Pytorch 1.8.0 [45]. For benchmarking tasks, we use the following ML models:

- ResNet-18 (18 layers, 11 million parameters) and ResNet-50 (50 layers, 23 million parameters) [30] with residual functions for image classification.
- Bert-Small [48] (66 million parameters) and Bert-Medium [55] (110 million parameters) for natural language processing.
- Atari break out game (50 million frames [34]) for reinforcement learning (RL).

In all experiments, we split the training dataset into smaller sets (up to 250 MB in size) and store them on AWS S3, which serves as our object store. We vary the memory allocated to training workers for each experiment based on the model size being evaluated. Unless otherwise noted, for hierarchical model synchronization in SMLT, we use the in-memory Redis key-value store hosted on AWS ECS, which serves as our parameter store. The AWS S3, AWS ECS, and the Lambda functions for workers are all deployed in the same AWS region, i.e., us-east-1.

## 5.2 Effectiveness of Hierarchical Model Synchronization

We evaluate the effectiveness of SMLT's proposed hierarchical model synchronization by comparing the communication time between SMLT and the two baselines, i.e., Siren [56] and Cirrus [22]. Figure 8 shows the communication time as a function of the number of workers (scale-out) per iteration for all 5 benchmarks, respectively. We observe that for all three systems the communication time increases linearly as the number of training workers increases. This increase is due to the fact that more workers will produce more gradients that need to be transferred from the workers to the aggregator via the external storage. However, the increase in overhead for SMLT is substantially lower than Siren and Cirrus, thanks to the hierarchical aggregation mechanism along with the fast parameter store, which together significantly reduce the communication overhead in a serverless environment.

For 2 representative benchmarks, we further breakdown the communication time into individual communication steps during each training iteration, as shown in Figure 7. While using SMLT, we denote *UL-Shard* as the time taken for the shard generator to split the gradients and upload them to the parameter store, *DL-Shard* as the time for the shard aggregator to download and aggregate the shards to form the aggregated shards, *UL-aggr* as the time taken for the shard aggregator to upload the aggregated shards, and *DL-grad* as the time for the global aggregator to download the aggregated shards (see Figure 5 for the detailed process description). For Siren and Cirrus, *UL-grad* refers to the time taken for each worker to upload gradients to the cloud storage, and *DL-grad* refers to the time taken for workers to download the parameters of all other workers for updating the model after every iteration. From Figure 7, one can see that, for both Siren and Cirrus, the main bottleneck often is the *DL-grad* step. On the other hand, SMLT's sharding approach for uploading and downloading gradients results in a significant reduction of the *DL-grad* overhead.

It is worth noting that, in the case of the RL-based Atari model, the size of the uploaded data is larger than the Resnet-50 model. The longer uploading time is due to the large simulation data shared by each worker after every iteration. We observe the impact of larger uploaded data in Figure 7[d-f]. In the case of Siren, this impact is more pronounced, reflecting the limitation of a centralized parameter server.

## 5.3 User-centric Deployments

We use the following two scenarios (exemplified in Section 3.2) to evaluate our user-centric deployment and execution approach. The scenario 1 is to, given a user-specified training time limit, optimize the monetary cost. The scenario 2 is to, given a user-specified monetary budget, minimize the training time. The results are shown in Figure 9 and Figure 10, respectively.

In Scenario 1, only SMLT meets the user-specified time limit as Siren and Cirrus do not consider such user requirements. SMLT also has the lowest cost, thanks to its optimized deployment and execution. Figure 9 also shows that SMLT would have been able to achieve the best accuracy with the most number of epochs at the lowest cost, if we had stopped training at the time limit (i.e., 1 hour). In Scenario 2, as shown in Figure 10, all three frameworks meet the user-specified monetary budget, although Siren and Cirrus achieve it by coincidence. SMLT achieves significantly lower training time compared to the other two frameworks, because of its optimizations to match the user-specified budget.

These results validate that SMLT can meet the user-specified time and budget requirements while optimizing the monetary cost and training time, respectively. This capability provides important practical benefits to ML practitioners compared to other frameworks that are oblivious to user requirements.

## 5.4 Dynamic Batching and Online Learning

To show how our task scheduler can exploit serverless computing's 'pay-as-you-go' pricing model using the Bayesian optimizer, we conduct experiments for two cases: *dynamic batching* and *online learning*. In these experiments, we compare SMLT with MLCD (a VM-based MLaaS platform [59]), LambdaML [33], and the IaaS setup as described in [33].

In the case of dynamic batching, the memory size and the number of workers can be scaled without restarting the worker functions. As shown in Figure 11a, the profiling overhead, and the cost associated with it, in SMLT is significantly lower than in MLCD. Although LambdaML may not perform as well as the IaaS setup, SMLT showcases that an ML training platform using serverless computing can provide a cost-effective and scalable solution.

In the case of online learning, due to the non-deterministic training time and the continuous resource provisioning, the cost for MLCD is higher than SMLT as shown in Figure 11b. Similarly, IaaS has high costs due to its continuously running, but at times idle, VM resources. On the other hand, LambdaML and SMLT showcase how an ML training platform can exploit the cost-effectiveness of serverless computing.

Note that, SMLT offers an adaptive approach for dynamic workload conditions, whereas LambdaML does not. To demonstrate this crucial difference, we conduct another experiment and compare SMLT's task scheduler to LambdaML (with a randomly-assigned, fixed resource allocation scheme). Figure 12a shows the training throughput (i.e., processed data points per second) over time, whereas Figures 12b and 12c show the corresponding number of workers and batch size. We assume that the user optimizes the resource allocation for the initial training configuration (i.e., batch size). Therefore, SMLT and LambdaML achieve similar training throughput initially. However, when the batch size changes, LambdaML's user-defined configuration observes sub-par training performance. In contrast, SMLT adjusts to the varying training parameter dynamically. We also observe the variation in throughput when the batch size changes are detected by the task scheduler via monitoring the workers. Cost-wise, SMLT also reduces the training cost by over 30% compared with LambdaML thanks to its dynamic resource allocation via the task scheduler.

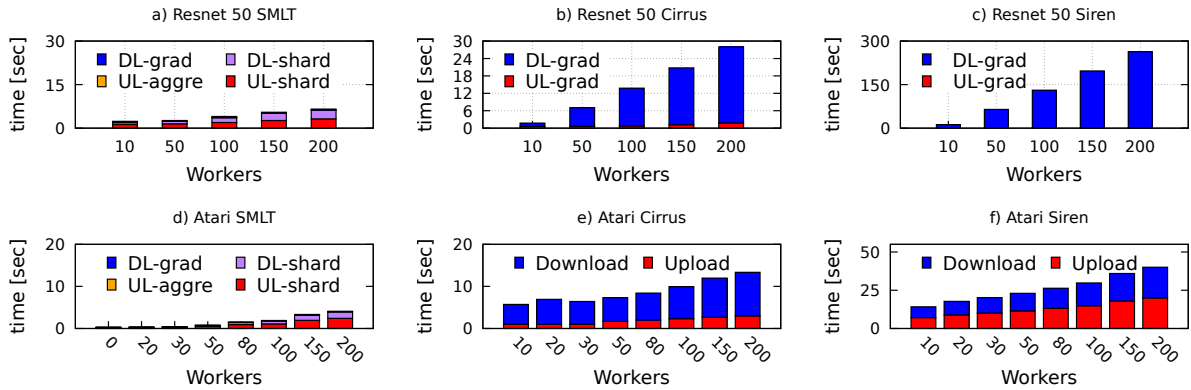

Figure 7: Communication time breakdown comparison of SMLT, Cirrus and Siren.

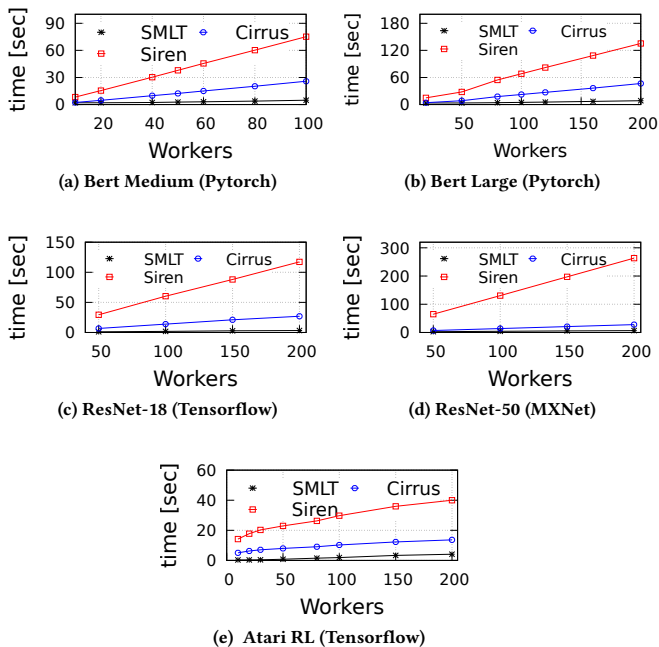

Figure 8: Per iteration communication time comparison between SMLT, Cirrus and Siren.

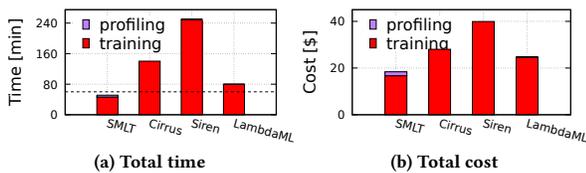

Figure 9: (Scenario 1) Minimize the monetary cost with a 1-hour training time limit for Bert-Medium with Pytorch. Note that, SMLT has profiling time and cost while other frameworks do not. For a fair comparison, we also demonstrate the profiling time and cost in SMLT.

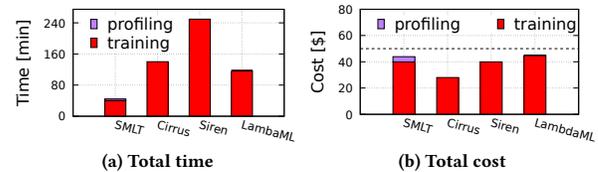

Figure 10: (Scenario 2) Minimize the training time with a $50 training cost budget for Bert-Medium with Pytorch. Note that, SMLT has profiling time and cost while other frameworks do not. For a fair comparison, we also demonstrate the profiling time and cost in SMLT.

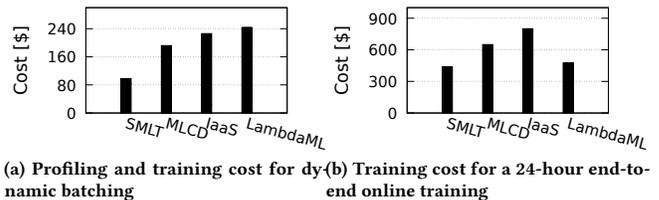

(a) Profiling and training cost for dynamic batching
(b) Training cost for a 24-hour end-to-end online training

Figure 11: Cost comparison of profiling and training via our Bayesian optimizer, for dynamic batching and online learning, using Resnet-50 with Pytorch. Experiments using Resnet-18 with Tensorflow produced similar results.

## 5.5 Neural Architecture Search

Neural architecture search (NAS) often deploys up to hundreds of training jobs in parallel or series to search for a well-performing model architecture. In such scenarios, each training job generally deploys a different model architecture as part of the exploration phase. As such, the amount of resources required is dynamic since it depends on the size of the model being deployed. We demonstrate the adaptive behaviour of SMLT for optimal resource allocation for the widely-used NAS framework called ENAS [46]. Figure 13a demonstrates the throughput comparison between SMLT and LambdaML. Similarly, Figures 13b and 13c display the corresponding changes in number of workers and model size over time, respectively. We assume that the user optimizes resource allocation for

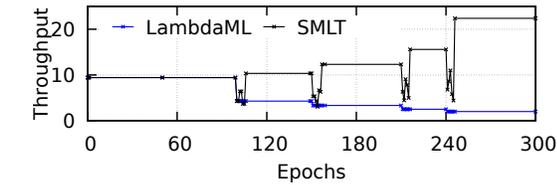
(a) Throughput comparison of SMLT and LambdaML

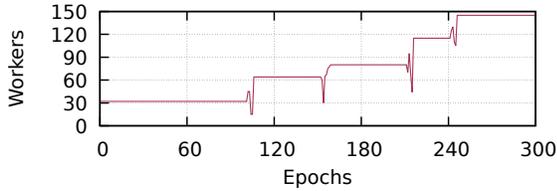
(b) Changes in the number of workers over time in SMLT

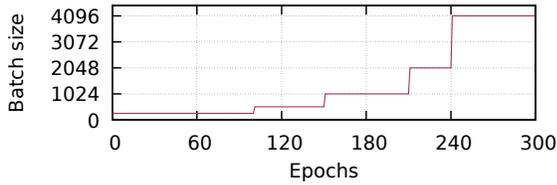
(c) Changes in batch size over time

Figure 12: Throughput comparison for dynamic batching.

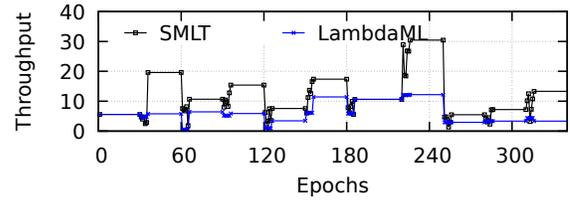
(a) Throughput Comparison of SMLT and LambdaML

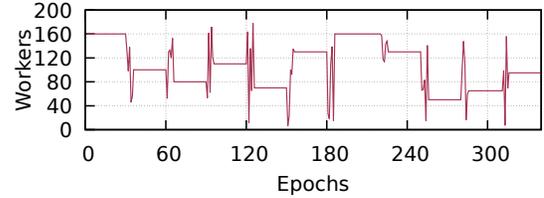
(b) Changes in the number of workers over time in SMLT

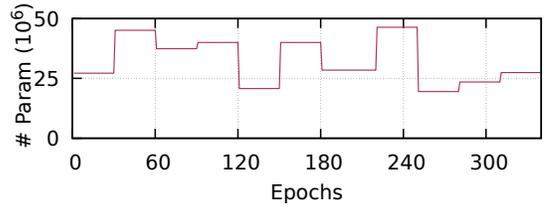
(c) Changes in model parameters over the course of ENAS.

Figure 13: Throughput comparison for ENAS.

LambdaML based on the first model. Therefore, both SMLT and LambadML achieve similar throughput at the beginning of the exploration. However, as the model size changes, SMLT optimizes the resource allocation accordingly, resulting in improved throughput. SMLT achieves 3× cost savings compared to LambdaML through its dynamic resource allocation scheme.

## 6 RELATED WORK

Like in any infrastructure-related problem, achieving a scalable and cost-effective solution becomes important for running ML workflows in the cloud. Earlier designs for MLaaS focus on addressing scalability [9, 63, 64], privacy [32, 54], and ease-of-use [53, 58], but not on cost-effectiveness. Major cloud providers, like Microsoft Azure [6] and AWS [13], have extended their existing statically-priced cloud platforms to commercial MLaaS without offering user-centric deployment guarantees. Similarly, recent academic efforts [37, 57, 62] utilize traditional cloud resources (e.g., VMs, containers) as their underlying platforms, but require extraneous management policies as pointed out in [35, 60].

Some previous systems attempt tackling the dynamic resource demands of ML tasks for specific use cases. DynamoML [25] proposes a set of online dynamic management techniques that perform scaling, job preemption, workload-aware scheduling and elastic GPU sharing for different parts of the ML development workflow consisting of modeling, training, and inference jobs. However, DynamoML is limited to non-serverless cloud settings, causing its resource allocation to be more coarse-grained, resulting in sub-par performance compared with a serverless setting. In addition, DynamoML requires prior knowledge of the workloads to be able to schedule the resource allocation effectively, making it unrealistic under workload settings, such as the NAS exploration phase without known resource requirements. Applications using complex ML techniques, such as the one described by Schuler et al. [50], use reinforcement learning that requires the training and tuning of the scheduler itself, adding an extra layer of overhead. Barista [19] uses a workload characterization and prediction mechanism to maintain SLOs with minimal cost. However, Barista's mechanisms are limited to the inference phase that has relatively less variation in resource dynamics than for the training. In contrast to all of these systems, SMLT does not require prior knowledge of the workloads because of its adaptive capability, does not require extra tuning of its scheduler supported by the Bayesian optimizer, and covers ML training phases that have more dynamic resource requirements.

## 7 CONCLUSION

We believe serverless plays a critical role in future machine learning routines. SMLT is the first fully-automated serverless framework for scalable and adaptive ML design and training. The key contributions of SMLT are to: 1) equip serverless platforms with an overarching view and dynamic adaptation capabilities for ML workflows, 2) offer user-centric deployment and execution of ML tasks on serverless platforms, and 3) provide an open-source end-to-end framework that provides fast gradient synchronization and addresses the key limitations of serverless ML platforms. Our extensive experimental evaluation demonstrates the effectiveness and robustness of SMLT, which outperforms the state-of-the-art approaches by up to 8× faster in training speed and 3× lower in monetary cost.

Ruichuan Chen2, Feng Yan1„